\documentclass[pre,twocolumn,superscriptaddress]{revtex4-1}

\usepackage{amsmath}    
\usepackage{graphicx}   
\usepackage{verbatim}   
\usepackage{color}      
\usepackage{hyperref}   
\usepackage{subcaption}

\begin{document}

\title{Collective sensing of $\beta$-cells generates the  metabolic code at optimal islet size}
\author{Dean Koro\v sak} 
\thanks{Corresponding author: dean.korosak@um.si}
\affiliation{University of Maribor, Faculty of Medicine, Institute for Physiology}
\affiliation{University of Maribor, Faculty of Civil Engineering, Transportation Engineering and Architecture}
\affiliation{Percipio, Ltd.}
\author{Marjan Slak Rupnik}
\thanks{Corresponding author: marjan.slakrupnik@muv.ac.at}
\affiliation{University of Maribor, Faculty of Medicine, Institute for Physiology}
\affiliation{Medical University of Vienna, Center for physiology and pharmacology, Institute for Physiology}

\date{\today}

\begin{abstract}
Major part of a pancreatic islet is composed of beta cells that secrete insulin, a key hormone regulating influx of nutrients into all cells in a vertebrate organism to support nutrition, housekeeping or energy storage. Beta cells constantly communicate with each other using both direct, short-range interactions through gap junctions, and paracrine long-range signaling. However, how these cell interactions shape collective sensing and cell behavior in islets that leads to insulin release is unknown. When stimulated by specific ligands, primarily glucose, beta cells collectively respond with expression of a series of transient Ca$^{2+}$ changes on several temporal scales. Here we analyze a set of Ca$^{2+}$ spike trains recorded in acute rodent pancreatic tissue slice under physiological conditions. We found strongly correlated states of co-spiking cells coexisting with mostly weak pairwise correlations widespread across the islet. Furthermore, the collective Ca$^{2+}$ spiking activity in islet shows on-off intermittency with scaling of spiking amplitudes, and stimulus dependent autoassociative memory features.    
We use a simple spin glass-like model for the functional network of a beta cell collective to describe these findings and
argue that Ca$^{2+}$ spike trains produced by collective sensing of beta cells constitute part of the islet metabolic code that regulates insulin release and limits the 
islet size. 

\end{abstract}

\maketitle

\section{Introduction}





Endocrine cells in vertebrates act both as coders and decoders of metabolic code~\cite{tomkins1975metabolic} that carries information from primary endocrine sensors to target tissues. In endocrine pancreas, energy-rich ligands provide a continuous input to a variety of specific receptor proteins on and in individual beta cells and initiate signaling events in and between these cells~\cite{henquin2009regulation}. In an oversimplified medical physiology textbook interpretation, glucose is transported into a beta cell through facilitated diffusion, is phosphorylated and converted within a metabolic black box to ATP, leading to closure of K$_{ATP}$ channels, cell membrane depolarization and activation of voltage-activated calcium channels (VACCs), followed by a rise in cytosolic Ca$^{2+}$ to a micromolar range and triggering of SNARE-dependent insulin release~\cite{ashcroft1989electrophysiology}. However, glucose may influence beta cells signaling through several additional routes. There may be alternative glucose entry routes, like for example active Na-glucose cotransport~\cite{tomita1976phlorizin, trautmann1987characterization}, alternative calcium release sites, like ryanodine~\cite{islam2002ryanodine} and IP$_3$ receptors~\cite{lang1999molecular} or glucose may directly activate the sweet taste receptor and initiate signaling~\cite{henquin2012pancreatic}, to name a few. Activation of a beta cell on a single cell level therefore likely involves triggering of a variety of elementary Ca$^{2+}$ events~\cite{berridge2000versatility}, which interfere in space and time into a unitary beta cell Ca$^{2+}$ response to support Ca$^{2+}$-dependent insulin release. This Ca$^{2+}$-dependent insulin release can be further modulated by activation of different protein phosphorylation/dephosphorylation patterns (PKA, PKC, Cdk5, etc)~\cite{skelin2011camp,mandic2011munc18} or other protein modifications~\cite{paulmann2009intracellular} to either reduce or increase the insulin output.

One of the important features of the sensory collectives is the optimization of the spatial relations between its elements to maximize the precision of sensing~\cite{fancher2017fundamental}. In islets of Langerhans, beta cells dwell as morphologically well defined cellulo-social collectives. These ovoid microorgans are typically not longer than 150 micrometers. The relatively small and constant pancreatic islet size is an intriguing feature in vertebrate biology. The size distribution of islets is comparable in humans, rodents and wider within different vertebrate species, irrespective of evident differences in overall body and pancreas size as well as total beta cell mass~\cite{dolenvsek2015structural, kim2009islet}. In mice, islet sizes range between 50 $\mu$m and 600 $\mu$m, with a median values below 150 $\mu$m~\cite{lamprianou2011high}. To accommodate differences in the body size, there is nearly a linear relationship between the total number of similarly sized islets and body mass across different vertebrate species~\cite{montanya2000linear,bouwens2005regulation}. However, why are islets so conserved in size is unknown. 

All beta cells within an islet collective represent a single functional unit, electrically and chemically coupled network, with gap junction proteins, Connexins 36 (Cx36)~\cite{bavamian2007islet}, for short-range interactions and with paracrine signalling~\cite{caicedo2013paracrine} for long-range interactions between cells. The unitary cell response in one beta cell influences the formation of similar responses in neighboring beta cells and contributes to coordination of a large number of beta cells~\cite{stovzer2013glucose, cigliola2013connexins}. Explorations of these functional beta cell networks, constructed from thresholded pairwise correlations of Ca$^{2+}$ imaging signals~\cite{stovzer2013functional,markovivc2015progressive, johnston2016beta}, showed that strongly correlated subsets of beta cell collective organize into modular, broad-scale networks with preferentially local correlations reaching up to 40 $\mu$m~\cite{markovivc2015progressive}, but understanding of mechanisms that lead to these strongly correlated networks states in beta cell populations is still lacking. We argue that beta cells sense, compute and respond to information as a collective, organized in a network similar to sensory neuron populations~\cite{schneidman2006weak,tkavcik2016information}, and not as a set of independent cells strongly coupled only when stimulation is high enough.

Here we analyze pairwise correlations of Ca$^{2+}$ spike trains (unitary beta cell responses on the shortest temporal scale) in beta cell collective recorded in fresh pancreatic tissues slice under changing glucose stimulation conditions (6 mM subtreshold - 8 mM stimulatory). We look at weak correlations between beta cells which we found to be widely spread across the islet~\cite{azhar2010correlations}. Guided by the use of statistical physics models in describing populations of neurons~\cite{schneidman2006weak,tkacik2009spin}, we use a simple spin glass model for Ca$^{2+}$ beta cells activity and show that it well captures the features observed in the measured data. In a way, we recognize this efficiency of simple models in both neuronal and endocrine cell collectives as one manifestation of the "beauty in function"~\cite{rasmussen1970cell}.

\section{Spin model of a $\beta$-cell collective}

Spin models have been borrowed from statistical physics to describe the functional behavior of large, highly interconnected systems like sensory neurons~\cite{schneidman2006weak,tkacik2009spin,tkavcik2014searching}, immune system~\cite{parisi1990simple}, protein interactions~\cite{bryngelson1987spin}, financial markets~\cite{bornholdt2001expectation,krawiecki2002volatility}, and social interactions between mammals~\cite{daniels2016quantifying, daniels2017control}.

The model of the islet consist of $N$ cells; at time $t$ each of the cells can be in one of two states, spiking or silent, represented by a spin variable $S_{i}(t) = \pm 1$, ($i = 1,...,N$).
The effective field $E_i$ of the i-th cell has two contributions: one from the cell interacting with all other cells with interaction strength $J_{ij}$, and one from external field $h_{i}$. We assume that interactions extend over the whole system. 
\begin{equation}
E_i(t) = h_{i}(t) + \sum_{j=1}^{N}J_{ij}S_j(t)
\end{equation}
At the next moment (t+1) each cell updates its state $S_{i}(t)$ with the probability $p$ to $S_{i}(t+1) = +1$ and with the probability $1-p$ to 
$S_{i}(t+1) = -1$. The probability $p$ depends on the effective field $E_i$ that the i-th cell senses:
\begin{equation}
p = \frac{1}{1+ \exp(-2E_i)}. 
\end{equation}
The interaction strength $J_{ij}$ is a fluctuating quantity with contributions from amplitude J common to all links and from the pairwise contributions with amplitude I ~\cite{krawiecki2002volatility}: $J_{ij} = J\lambda(t) + I\nu_{ij}(t)$. 
Here are the fluctuations $\lambda(t)$  and $\nu_{ij}(t)$ random variables uniformly distributed in the interval 
$[-1,1]$. The external field $h_{i}(t) = h\eta_{ij}$ is also a random variable, uniformly distributed in the interval 
$[h_{min},h_{max}]$. In the mean-field approximation the average state of the system
$m(t) = \frac{1}{N}\sum_{j}S_j$, evolves with time according to~\cite{krawiecki2002volatility}: 
\begin{equation}
m(t+1) = \tanh(J\lambda(t)m(t) +h_{mf}(t)),
\label{mfeq}
\end{equation}
where we set $h_{mf} = (h/N)\eta(t)$. For the computations here, we estimate the boundaries of the external field interval $h_{max(min)}$
from non-interacting mean field model ($J=0$) corresponds to non-stimulating
glucose concentration (6 mM), so that $h_{mean} = \tanh^{-1}(m_{mean})$ and $h_{max(min)} = h_{mean} \pm \Delta h$. 
In our models we set $I = 2J$ for the pairwise interaction amplitude. 

\section{Results}

The functional multicellular imaging (fMCI) records a full temporal activity trace for every cell in an optical plane of an islet from which meaningful quantitative statements about the dynamics of unitary Ca$^{2+}$ responses and information flow in the beta cell collective are possible~\cite{stovzer2013glucose, dolenvsek2013relationship}. Briefly, after the stimulation with increased glucose level, first asynchronous Ca$^{2+}$ transients appear, followed by a sustained plateau phase with oscillations on different temporal scales, from slow oscillations (100-200 secs) to trains of fastest Ca$^{2+}$ spikes (1-2 secs). As the relation between the rate of insulin release and cytosolic Ca$^{2+}$ activity shows saturation kinetics with high cooperativity~\cite{skelin2011camp}, the insulin release probability is significantly increased during these Ca$^{2+}$ spikes. 

Initially, fMCI has been done at the glucose concentrations much higher than those at which beta cells usually operate. The main reason for this was to ensure comparability of the results with the mainstream research in the field using mostly biochemical approaches. At 16 mM glucose, a collective of beta cells responds in a fast, synchronized, and step-like manner. Therefore the first interpretation has been that gap junction coupling between neighboring beta cells presents a major driving force for the beta cell activation and inhibitory dynamics~\cite{markovivc2015progressive, hraha2014phase}. Accordingly, the removal of Cx36 proteins does cause hyperinsulinemia at resting glucose levels and blunted responses to stimulatory glucose concentration~\cite{speier2007cx36}. Since beta cells in fresh pancreatic tissue slices are sensitive to physiological concentration of glucose (6-9 mM)~\cite{speier2003novel}, we here focused on this less explored concentration range. We looked at the spiking part of the Ca$^{2+}$ imaging signals for which it has been previously shown to contain enough information to allow reconstruction of functional cell networks~\cite{stetter2012model}.

\begin{figure}[h!]
\centering
\includegraphics[scale=0.2]{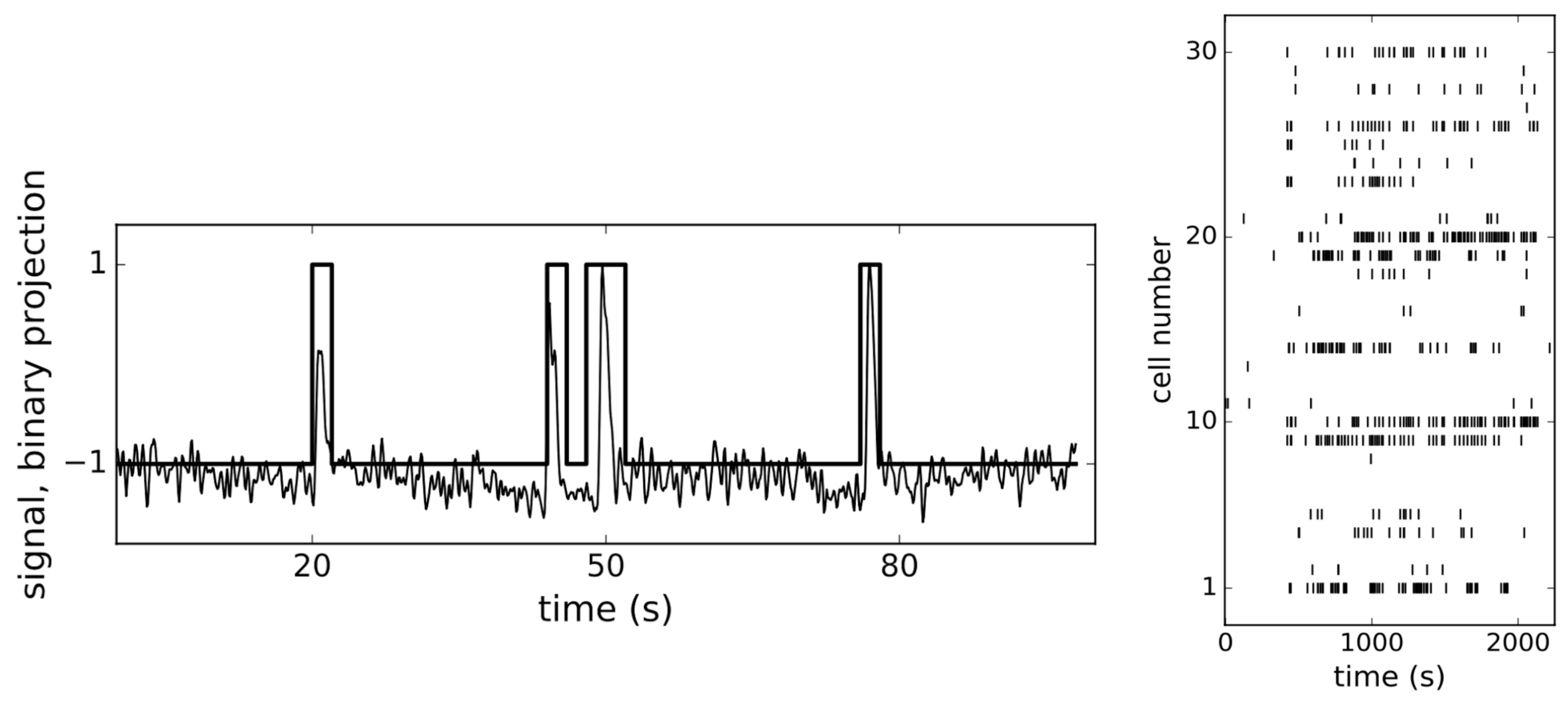}
\caption{\label{fig:fig1} (left) A Ca$^{2+}$ trace showing a short train of spikes after ensemble empirical mode decomposition with overlaid binary form with 2 s wide bins. 
(right) spin raster plot of 30 randomly picked beta cells}
\end{figure}

For the present analysis we used a dataset of individual Ca$^{2+}$-dependent events from $N=188$ ROIs with known positions from the central part of the fresh rodent pancreatic oval shaped islet (370 um in length and 200 um wide), representing beta cells, recorded with fMCI technique at 10 Hz over period of 40 minutes.
During the recording the glucose concentration in the solution filling the recording chamber has been increased from 6 mM to 8 mM, reaching equilibration at around 200 s after the start of the experiment, and then decreased to initial concentration near the end of experiment at around 2000 s. We applied ensemble empirical mode decomposition~\cite{luukko2016introducing} on recorded traces to isolate the Ca$^{2+}$ spiking component of the signal. Finally, we binarized the signals using 2 s wide bins (Fig. 1, left panek) and obtained binary spike trains $S_{j}(t)\pm 1$, ($j=1...N$), of beta cells' Ca$^{2+}$ activity, each cell represented as a spin. An example of spiking dynamics of 30 randomly chosen spins is shown as a raster plot in the right panel of Fig. 1.

Similarly to previous work in neuronal populations in vertebrate retina, we used the principle of maximum entropy mostly based on pairwise correlation between cells and which has been successfully used in predicting spiking patterns in cell populations~\cite{schneidman2006weak,tkavcik2014searching,tkacik2009spin,ferrari2017random}. It may seem surprising that models with first and second-order correlation structure work not only when the cell activity is very sparse so the correlations could be described by perturbation theory~\cite{roudi2009pairwise}, but can reproduce the statistics of multiple co-spiking activity~\cite{barton2013ising,merchan2016sufficiency,ferrari2017random}. 
We computed truncated correlations
\begin{equation}
c(i,j) = \langle S_iS_j\rangle - \langle S_i\rangle\langle S_j\rangle
\end{equation}
for all pairs of cells. The pairwise correlations found are mostly weak with the distribution shown in Fig. 2 (left panel), but they extend widely over the distances up to 170 $\mu$m across the islet, which is larger than an average vertebrate islet size (Fig2., right panel). At distances larger than 170 $\mu$m the correlations decrease sharply towards zero. Such weak and long-ranging pairwise correlations could be the root of criticality and of strongly correlated
network states in biological systems~\cite{mora2011biological,schneidman2006weak,azhar2010correlations,tkavcik2015thermodynamics}. 

\begin{figure}[h!]
\centering
\includegraphics[scale=0.3]{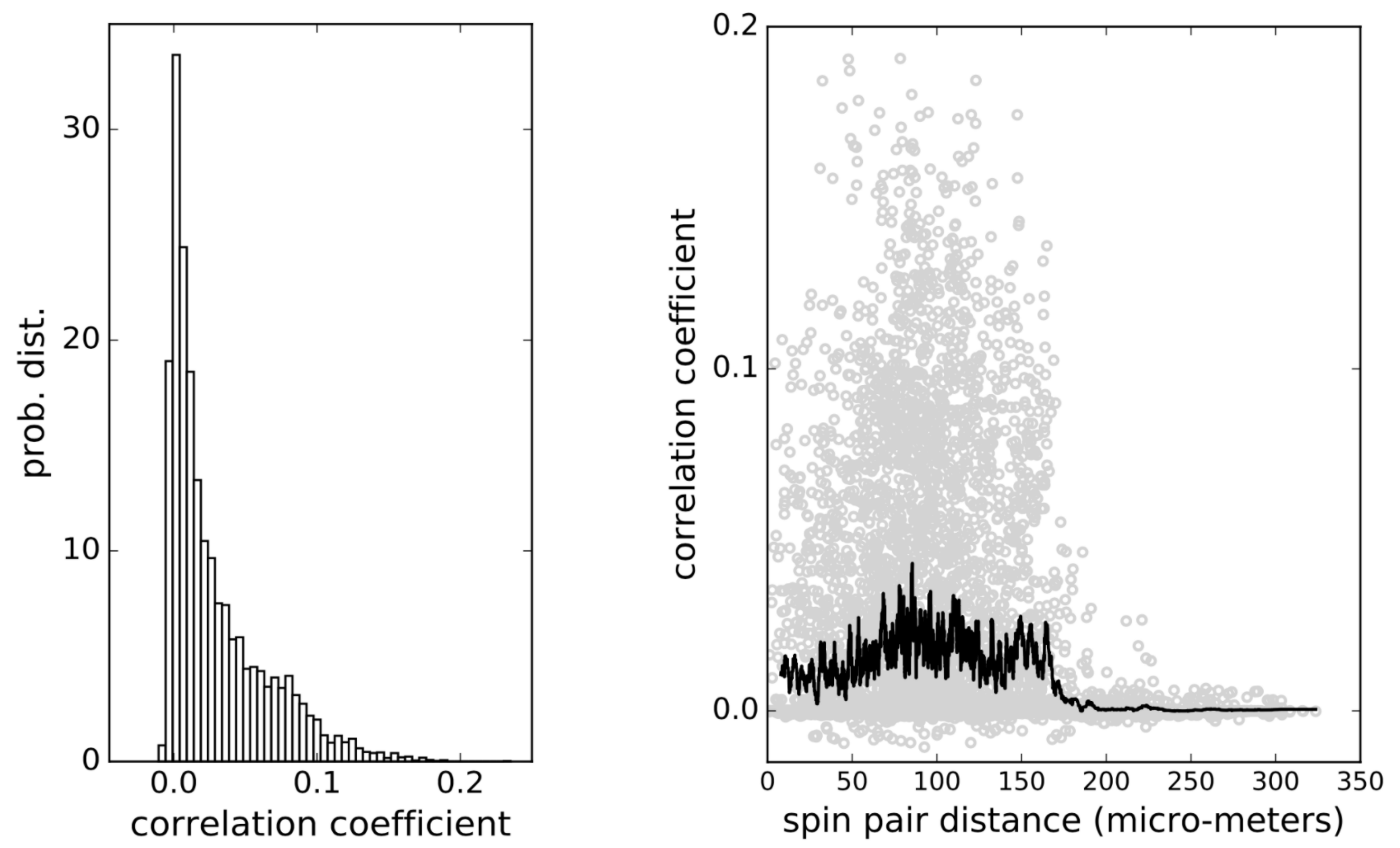}
\caption{\label{fig:fig1a} (left) Distribution of pairwise correlations of beta cell collective computed from Ca$^{2+}$ imaging spiking signals. (right) Pair correlations distribution over distance. Weak correlations extend over the whole system up to 170 $\mu$m. Black line shows the average values of correlations at particular cell-cell distances.}
\end{figure}

To check for the existence of strongly correlated states in weakly correlated beta cell collective we computed probability distributions $P_{N}(K)$ of $K$ simultaneously spiking cells in groups of $N=10,20,30$ cells. While the $P_{N}(K)$ of randomly reshuffled spike trains expectedly follows Poisson distribution (left panel in Fig. 3, black crosses and dashed line for $N=10$ spins), the observed co-spiking probabilities are orders of magnitude higher (diamonds in left panel of Fig. 3 for $N=10$ spins) than corresponding probabilities in groups of independent spins. The statistics of these co-spiking events were described by an exponential distribution~\cite{schneidman2006weak}, by finding the effective potential~\cite{tkavcik2013simplest,tkavcik2014searching} matching the observed $P_{N}(K)$ and adding it to the hamiltonian of pairwise maximum 
entropy model, or by using beta-binomial distribution~\cite{nonnenmacher2016signatures} $P_{N}(K) = C(N,K)B(\alpha+K,\beta+N-K)/B(\alpha,\beta)$ where $C(N,K)$ is binomial coefficient and $B(\alpha,\beta)$ is the beta function.  

\begin{figure}[h!]
\centering
\includegraphics[scale=0.3]{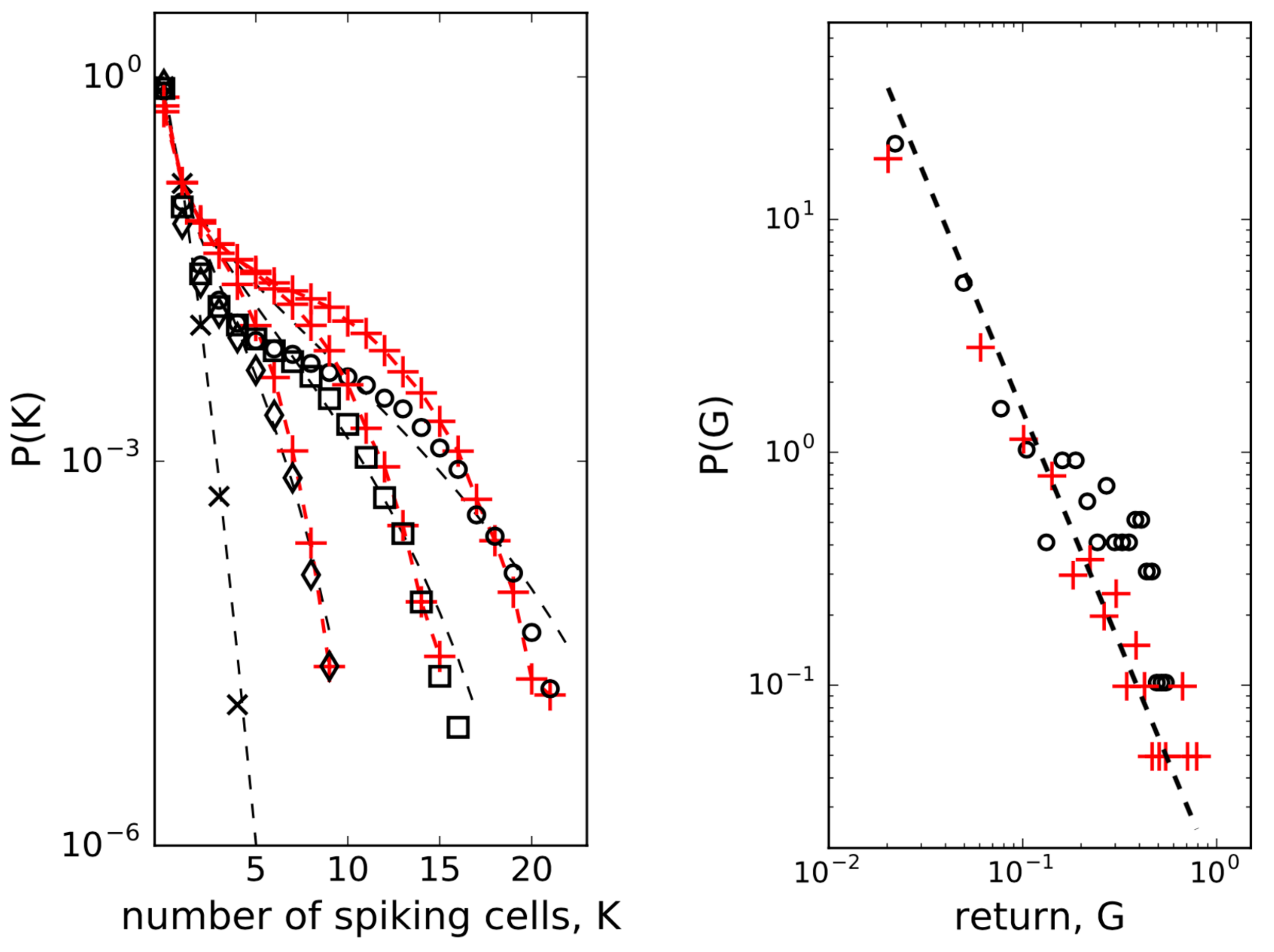}
\caption{\label{fig:fig3} (left) Probability distributions of $K$ cells among $N$ spiking simultaneously. Randomly shuffled spike trains (black crosses, $N=10$) with dashed line - Poisson distribution; $N=10$ (diamonds), $N=20$ (squares), $N=30$ (open dots); model (red pluses + red dashed line, $J=2.0$, $N=200$, see main text), 
beta-binomial model~\cite{nonnenmacher2016signatures} (black dashed line, $\alpha=0.38$, $\beta=11.0$);(right) Scaling of mean field return: open dots - data, red pluses - mean field approximation from the spin model of $\beta$-cells computed with $J=2.0$, $h_{mf} = (h/N)\eta(t)$. Dashed line $P(G) \sim G^{-2.0}$}
\end{figure}

We next run the spin model of 200 $\beta$-cells and then sampled the computed spike trains to obtain $P_{N}(K)$ from the model 
for $N=10,20,30$. Despite its simple structure, the model matches order of magnitude of the observed $P_{N}(K)$ well when we set the interaction strength at $J=2.0$, as shown in the left panel of Fig. 3 (red pluses and red dashed line), particularly for larger $K$ values. For comparison, we also show how the beta-binomial model fits to the observed data using the parameters $\alpha=0.38$, $\beta=11.0$ in all $N=10,20,30$ cases. These values are also close to the best-fitting parameters ($\alpha=0.38$, $\beta=12.35$)  to the simulated and observer correlated neural population activity data as reported in ~\cite{nonnenmacher2016signatures}.

The microscopic model of interacting spins with interactions randomly varying in time~\cite{krawiecki2002volatility}, adopted here to describe interacting $\beta$-cell collective, exhibits scaling of price fluctuations~\cite{bornholdt2001expectation} observed in financial markets~\cite{gopikrishnan1999scaling} and on-off intermittency with attractor bubbling dynamics of average price~\cite{krawiecki2002volatility}. Following this idea, we looked at the logarithmic return of average state of $\beta$-cell collective at time $t$~\cite{bornholdt2001expectation}:
$G(t)=\log(m(t))-\log(m(t-1))$. As presented in the right panel of Fig. 3, the distribution $P(G)$ (of positive $G$ values) can indeed be approximated with a scaling law: $P(G)\sim G^{-\gamma}$ with $\gamma=2.0$. Computing the average state with the eq.(\ref{mfeq}) of the model, we can reproduce the observed distribution by setting on the interaction strength to $J=2.0$ at $t_{on}=400$ s and off to $J=0$ at $t_{off}=2300$ s. The amplitude of the interaction $J$ is consistent with the computation of the co-spiking probability.   

\begin{figure}[h!]
\centering
\includegraphics[scale=0.35]{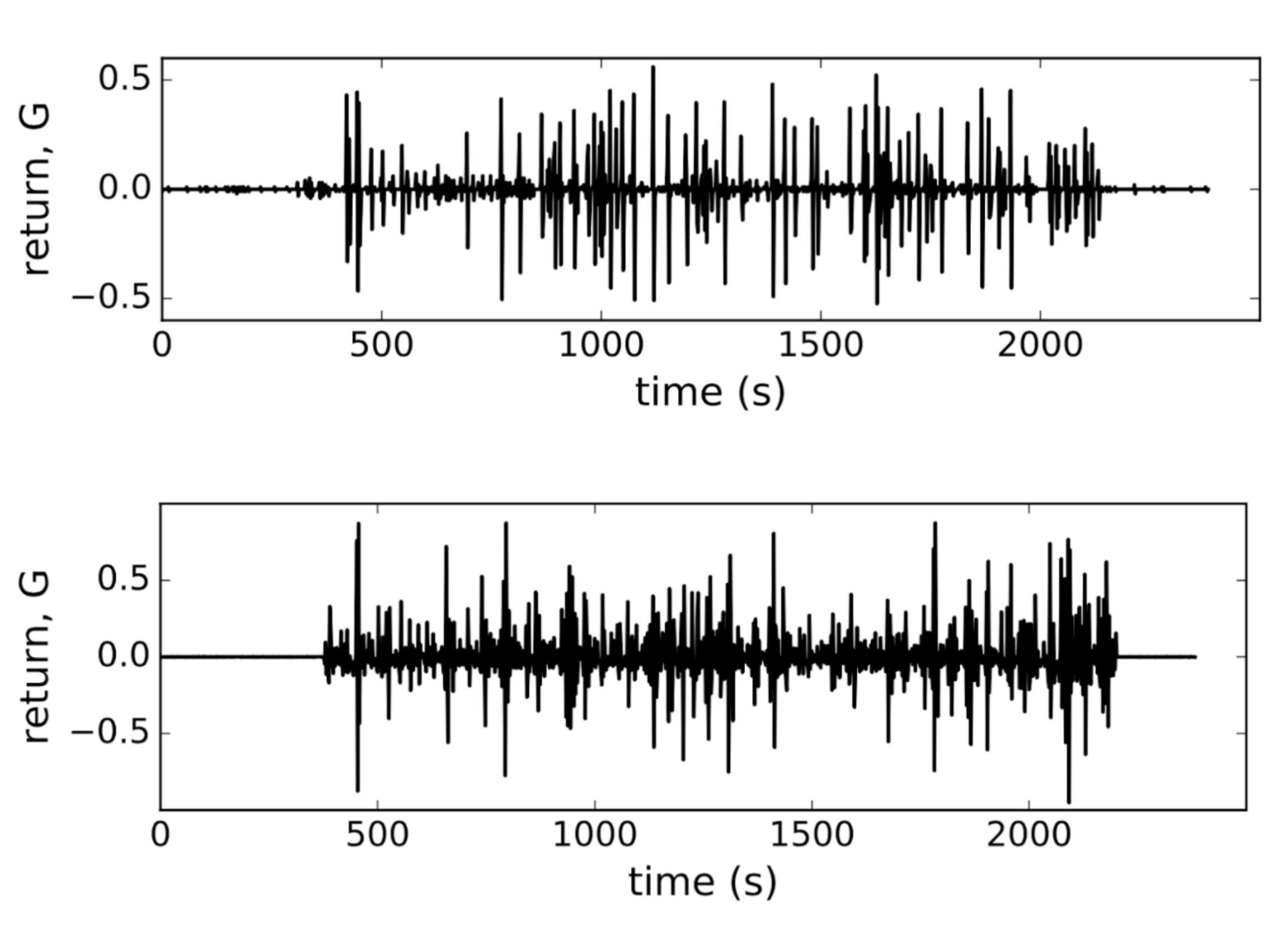}
\caption{\label{fig:fig2} (upper) Observed logarithmic return of the average state of  of $\beta$-cell collective $G(t)$, (lower) logarithmic return of the average state computed from the model with $J=2.0$ for $t_{on}<t<t_{off}$.}
\end{figure}

In Fig. 4 we show the plots of both, observed and computed, returns of average state of interacting $\beta$-cells for comparison. The glucose concentration 
was changed during the experiment in a stepwise manner: from 6-8 mM at the beginning and back to 6 mM near the end of recording period. 
The effect of both changes is nicely visible in the $G(t)$ plot (upper panel, Fig. 4) where the on-off intermittent dynamics of the average state starts around
$t_{on} = 400$ s and lasts until around $t_{off} = 2300$ s in the experiment. Both observed events are delayed with respect to the times of glucose concentration change due to the asynchronous Ca$^{2+}$ transients~\cite{stovzer2013glucose}. We expect that the response of $\beta$-cell collective to the stimulus increase must be visible in the variance of average state $\textrm{Var}(m)$ which is in Ising-like model we are using here equal to susceptibility of the system 
$\chi =\textrm{Var}(m)=\langle m^2\rangle-\langle m\rangle ^2$. 
In upper panel of Fig. 5 (open black dots) we show the plot of susceptibility as a function of recording time, focusing around the transition to increased glucose 
concentration during the experiment. There is a sharp increase of susceptibility at around $t_{on}$, the same time the on-off intermittency starts to appear in $G(t)$. Using mean field approximation of the spin model eq.(\ref{mfeq}) for computation of susceptibility (averaged over many runs) and setting $J = 0$ for $t < t_{on}$ and $J = 2.0$ for $t > t_{on}$ we can well describe the observed evolution of susceptibility and capture the rapid onset of increased sensibility of the islet (red line in upper part of Fig. 5). 

\begin{figure}[h!]
\centering
\includegraphics[scale=0.3]{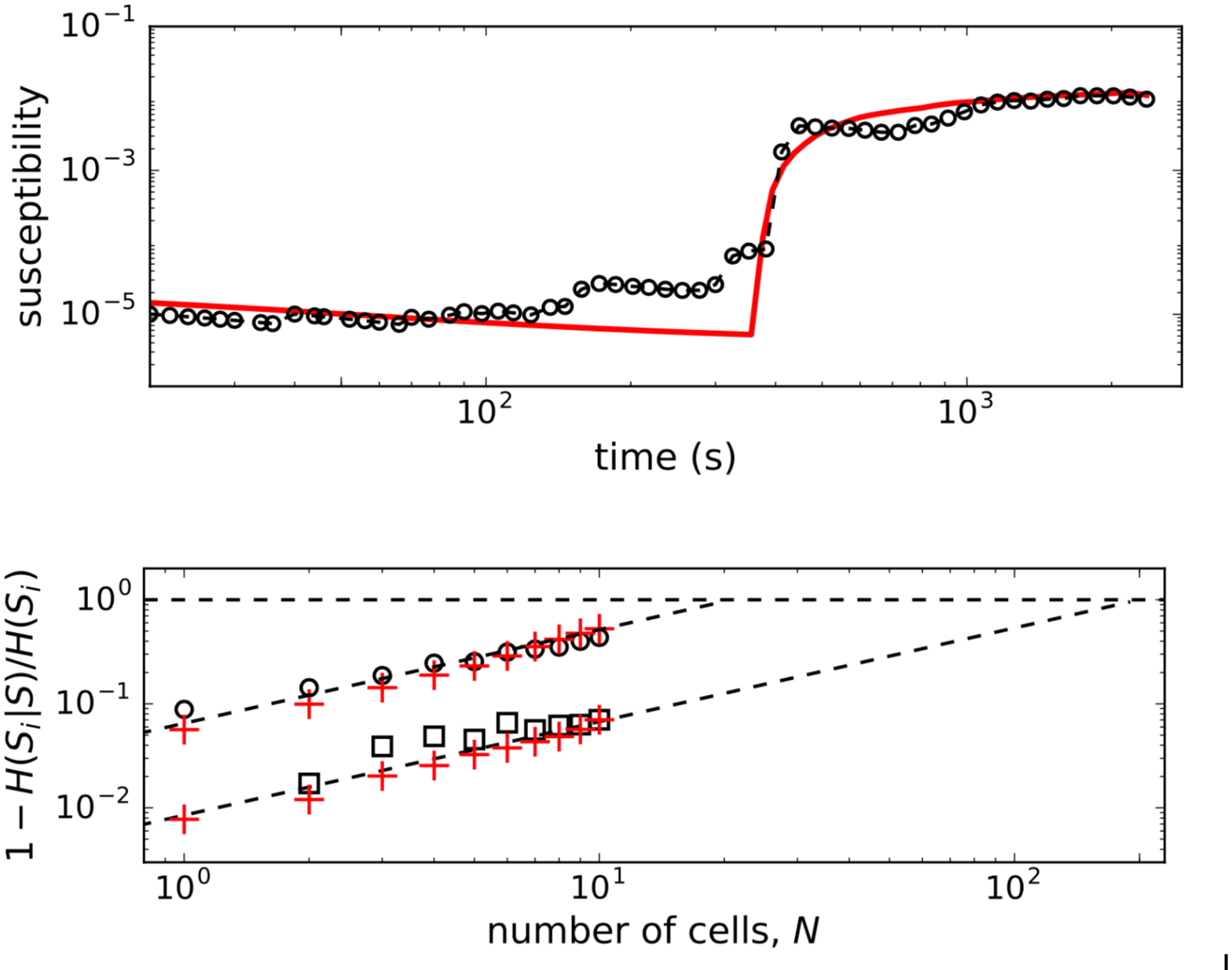}
\caption{\label{fig:fig4} (upper panel) Susceptibility of $\beta$-cell collective around transition to stimulatory glucose level. Open dots are the experimental data, red line shows the result of the mean field computations with  $J = 2.0$ onset at $t=t_{on}$;(lower panel) Normalized conditional entropy. Open dots are experimental data at 8 mM glc, open squares at 6 mM glc. Red pluses show the results of the spin model computations with $N_{spins}$=200 spins, and the parameters: $h_{min}$=-2.65, $h_{max}$ = -1.65, $J=2.0$ for the upper, and $J=0$ for lower the lower part}
\end{figure}

Pairwise correlation structure enables error-correction features of population coding in neural systems~\cite{schneidman2006weak}.  
To check for memory-like or error-correcting properties in islets, we use the conditional entropy $H(S_i|S)$, the measure for the information we need to determine the state of $N$-th cell (i.e. spiking or not) if we know the states of $N-1$ cells ($S={S_j\neq i}$) in a group of $N$ cells. If the state of the $N$-th cell is completely determined by other $N-1$ cells, the conditional entropy is zero $H(S_i|S) = 0$ and the error correction is perfect. When $S_j$ are independent random states, the conditional entropy equals the entropy of the $N$-th cell $H(S_i)$. 

We computed the quantity $1-H(S_i|S)/H(S_i)$ (normalized mutual information) as a function of number of cells (for small groups of cells) and extrapolate the trend towards the limit $H(S_i|S) = 0$ that determines the critical number of cells, $N_c$, needed to predict the state of another cell. As seen in the lower panel of Fig. 5, the predictability is a glucose-dependent parameter. With non-stimulatory glucose concentration, the complete set of data is required for predictions, whereas at 8 mM glucose we find that order 
of magnitude smaller number of measured cells are needed to predict the states of other cells.


\section{Discussion}

Pancreatic beta cell continuously intercepts a variety of energy-rich or signaling ligands using the whole spectrum of specific receptors on the cell membrane, as well as in metabolic and signaling pathways within the cell. The cell converts these signals into a binary cellular code, for example a trains of Ca$^{2+}$ spikes, which drive insulin release that fits current physiological needs of the body. This way, already a single cell can sense its chemical environment with extraordinary, often diffusion limited precision~\cite{bialek2005physical}, however, judging by their heterogeneous secretory behavior in cell culture, the precision of sensing among the individual beta cells is quite diverse~\cite{hiriart1991functional}. Recent experimental evidence and modeling have shown that cell collectives sense better compared to an individual cell. The precise mechanism of this collective sensing improvement depends on cell-cell communication type, which can be short-range with direct cell contacts or long-range with paracrine signaling~\cite{fancher2017fundamental}. Furthermore, also long-range interaction have its finite reach which can poise a limit to the cell collective size and therefore determines its optimal as well as maximal size. As mentioned in the Introduction, it is intriguing how well conserved is the pancreatic islet size in vertebrates of dramatically different body dimensions~\cite{montanya2000linear}. In a single vertebrate organism the size of the islets can be bigger that 150 um, but functional studies revealed that the islets bigger than 200 um secrete 50\%  less insulin after glucose stimulation~\cite{fujita2011large}. These functional differences between small and large islets have been partially attributed to diffusion barriers for oxygenation and nutrition, limiting the survival of core beta cells in bigger islets after isolation. However, reducing these diffusion barriers had no influence on insulin secretory capacity~\cite{williams2010reduction} suggesting other factors, like diffusion of paracrine signaling molecules~\cite{caicedo2013paracrine} could limit the collective beta cell function in bigger islets. This dominance of a long-range information flow, likely limited to some physical constraints, has indicated the use of the mathematical equivalency with spin glass-like systems~\cite{tkavcik2016information}. 

We strongly believe that advanced complex network analysis based on strong short-range correlations can continue to provide valuable information regarding the beta cell network topologies, network on network interactions and describe the functional heterogeneity of individual beta cells~\cite{markovivc2015progressive,johnston2016beta,gosak2015multilayer}. However, the main goal of the present study was to determine the influence of weak long-range correlations between pairs of beta cells on the probability of activation of single beta cells. As has recently been shown that it is sufficient to use pairwise correlations to fully quantitatively describe the collective behavior of cell collectives~\cite{merchan2016sufficiency}. The typically small values of pairwise correlation coefficients with the median values below 0.02, would intuitively be ignored and beta cells described as if they act independently. However, it has been shown that in larger populations of cells this assumption fails completely~\cite{schneidman2006weak}. In fact, at physiological stimulatory glucose levels between 6 and 9 mM, beta cell collectives are entirely dominated by weak average pairwise correlations (Fig. 2). Nevertheless, this is the glucose concentration range, where beta cells are most responsive to the nutrient to, as a collective, compute their activity state and pulsatile insulin release, and to meet the organismal needs between the environmental and behavioral extremes of food shortage and excess~\cite{schmitz2008high}? 

Based on the range of the calculated weak pairwise correlations of up to 170 um (Fig. 2), we predict that beta cells collective falls into a category of sparse packed tissues with dominant paracrine interactions and that cell-cell distances contribute to optimal sensing and functional response in creating the metabolic code governing the release of insulin. It remains unclear whether and how the position of beta cells within an islet is controllable. As many other cells, beta cells are polarized and possess a primary cilium~\cite{gan2017cell}, which should have a primary role in sensory function, i.e. insulin sensing in paracrine signaling~\cite{douganer2016autocrine}, and not in cell motility. It is quite interesting though, that the ciliopathies are highly associated with reduced beta cell function and increased susceptibility to diabetes mellitus~\cite{gerdes2014ciliary}. Future experiments are required to test for the possible motility of beta cells within the islet to adopt an optimal separation of key sensitive beta cells. To further extrapolate the collective sensing idea, it is also possible that the diffuse arrangement of a collective of islets within different parts of pancreas, which are exposed to different vascular inputs~\cite{dolenvsek2015structural}, serves to optimize nutrient sensing experience, yet on a higher organizational level,  providing a topological information regarding the nutrient levels in different parts of the gastrointestinal tract. The nature and level of interactions between individual islets in the pancreas are currently also unknown.

As in retinal neuron networks, beta cells encode information about the presence of energy-rich nutrients into sequences of intermittent Ca$^{2+}$ spikes. In a natural setting of sensory neural networks with stimuli derived from a space with very high dimensionality the coding seems challenging and interpretations require some strong assumptions~\cite{tkavcik2014searching}. We currently do not understand the input dimensionality of a typical ligand mixture around the beta cells, we simply assume it is not high. As in retinal networks~\cite{tkavcik2014searching,schneidman2006weak}, the predictability regarding the functional state of individual beta cells is defined by the network and not the chemical environment. This suggests that the sensory information at physiological glucose levels is substantially redundant. It is likely that the nutrient mixture presents a noisy challenge for the information transfer which is typical for biological system. But why study the insulin release pattern or the metabolic code? The beta cell network possess associative or error-correcting properties (Fig. 5), so this idea from the sensory neuron networks can be generalized also to populations of endocrine cells~\cite{schneidman2006weak}, which may again influence the optimal islet size and suggest the presence of functional subunits within the islet that could adapt, for example, to changing environment in a dynamic fashion. Furthermore, error-correction properties are glucose dependent and can be physiologically modulated (Fig. 5). The trains of Ca$^{2+}$ spikes at constant glucose stimulation (8 mM) are inhomogeneous, display on-off intermittency (Fig. 4) and scaling of log returns of 
average state (Fig. 3) analog to models of financial time series~\cite{krawiecki2002volatility}. Also here, the sources of stochasticity in an islet collective are various. On one hand, the beta cells make decisions on activation under the influence of the external environment and other beta cells. Second, also the time-dependent interaction strength among beta cells is random, which could reflect their socio-cellular communication network and indicate that the external environment can be sensed differently between different beta cells in an islet.

Biological systems seem to poise themselves at criticality, with a major advantage of enhanced reactivity to external perturbations~\cite{mora2011biological}. Often a limited number of individual functional entities, cell or groups of cells as found in pancreatic islets, appeared to be limiting to address criticality. However, it has been recently demonstrated that even in biological systems with small number of interacting entities one can operationally define criticality and observe changes in robustness and sensitivity of adaptive collective behavior~\cite{daniels2017control}. Our results suggest that beta cells collective within the islet sits near its critical point and we could determine the susceptibility in the islet. Stimulatory glucose concentration (8 mM) has been decreasing distance to criticality by increasing sensitivity (Fig. 5). Smaller distance to criticality at unphysiologically high glucose levels has its possible adverse consequences in a phenomenon called critical slowing down as the system takes more and more time to relax as it comes nearer to the critical point~\cite{mora2011biological}. Our preliminary results show that at very strong stimulation (i.e. glucose levels above 12 mM) the whole system freezes into a certain state where short-term interaction take over enabling global phenomena within the islets, e.g. Ca$^{2+}$ waves~\cite{stovzer2013functional} requiring progressively longer periods to relax to baseline with increasing glucose concentrations. 

The exact nature of the relation between the islet size and collective sensing in pancreatic islets is not clear. In the pathogeneses of different types of diabetes mellitus, the islet size is an important parameter. Until recently it has been thought that in type 1 diabetes mellitus, insulin release is no longer functional. We now know  that even in type 1 diabetic patients small and functional collectives of beta cells persist in the pancreata of these patients even decades after the diagnosis~\cite{faustman2014were}. 
On the other hand, the beta cells mass in an islet can be increased in type 2 diabetic patients in the initial phases after the diagnosis ~\cite{rahier2008pancreatic} or in animal models~\cite{daraio2017snap} and can only be reduced in the later phases~\cite{rahier2008pancreatic}. It remains to be established what are the relations between the reduced or increased insulin release, changed islet size and therefore changed circumstances for paracrine signaling in disturbed collective nutrient sensing and during the aforementioned pathogeneses of diabetes mellitus.

\subsection*{Acknowledgements}

The authors acknowledge the financial support from the Slovenian Research Agency (research core funding, No. P3-0396), as well as research project, No. N3-0048).

\bibliography{refs}{}
\bibliographystyle{apsrev4-1}

\end{document}